\documentclass{PoS}

\def\ga{\mathrel{\mathchoice {\vcenter{\offinterlineskip\halign{\hfil
$\displaystyle##$\hfil\cr>\cr\sim\cr}}}
{\vcenter{\offinterlineskip\halign{\hfil$\textstyle##$\hfil\cr>\cr\sim\cr}}}
{\vcenter{\offinterlineskip\halign{\hfil$\scriptstyle##$\hfil\cr>\cr\sim\cr}}}
{\vcenter{\offinterlineskip\halign{\hfil$\scriptscriptstyle##$\hfil\cr>\cr
\sim\cr}}}}}

\title{Ultra-High Energy Cosmic Ray and Neutrino Observations}

\ShortTitle{UHECR and $\nu$-Observations}

\author{\speaker{Karl-Heinz Kampert}\thanks{Invited paper given at the EPS Conference on High Energy Physics, 2009}\\
        Bergische Universit\"at Wuppertal, Department of Physics, 42119 Wuppertal (Germany)\\
        E-mail: \email{kampert@uni-wuppertal.de}}

\abstract{Recent measurements of ultra-high energy cosmic rays and neutrinos are reviewed. With several new large scale observatories nearing completion or becoming fully operational only very recently, a large body of high quality and high statistics data is growing up now. Already these first data have started to open up a new window to the high energy Universe giving us first direct clues about the origin of the most energetic particles with energies of about $10^{20}$~eV as well as about their interactions from extragalactic sources to Earth. Also, for the first time full sky views of high energy neutrinos have become available with neutrino telescopes operating on either Hemisphere. While a ``smoking gun'' is still missing on galactic sources of cosmic rays, constraining upper limits to neutrino fluxes from various source candidates are reported. Thus, future neutrino telescopes, such as {\sc KM3NeT} in the Mediterranean should aim at volumes significantly larger than one cubic kilometer. Besides seeking the sources of galactic and extragalactic cosmic rays, the new generation of cosmic ray and neutrino observatories touches a wide range of scientific issues and they have already provided important results on tests of fundamental physics.}

\FullConference{European Physical Society Europhysics Conference on High Energy Physics,
EPS-HEP 2009,\\
		 July 16 - 22 2009\\
		 Krakow, Poland}

\begin{document}

\section{Cosmic Rays and Neutrinos: Science Case}

Understanding the origin of the highest energy cosmic rays is one
of the most pressing questions of astroparticle physics. Cosmic
rays (CRs) with energies exceeding $10^{20}$ eV have been observed for more than 40 years (see e.g.\ \cite{Nagano-Watson,Bluemer-09}) but due to their low flux only some ten events of such high energies could be detected up to recently.  There are no generally accepted source candidates known to be able to produce particles of such extreme energies \cite{Hillas84}.
The requirements are not easily met, which has stimulated the production of a large number of creative papers. Moreover, there should be a steepening in the energy spectrum near $10^{20}$ eV due to the interaction of CRs with the microwave background radiation (CMB).  This Greisen-Zatsepin-Kuzmin (GZK) effect \cite{GZK} severely limits the horizon from which particles in excess of $\sim 6\cdot10^{19}$ eV can be observed.  For example, the sources of protons observed with $E\ge 10^{20}$ eV need to be within a distance of less than 50 Mpc \cite{Harari-06}.

Ironically, tackling the problem of CR origin from the upper end of the energy spectrum, where their measurements are a most challenging experimental task, appears to be the most promising approach. The reason for this is twofold: at the highest energy ($E \ga 5 \cdot 10^{19}$~eV), deflections in galactic and extragalactic magnetic fields are considered small enough to allow performing CR astronomy, i.e.\ the most energetic particles should point back to the direction of their sources. Secondly, the GZK-effect suppresses particles from distances larger than 50-100~Mpc, so that it acts as a filter to nearby sources minimizing directional ambiguities arising from too many sources. 

Besides astrophysics, there is also a particle physics interest
in studying this energy regime. This is because CRs give access to elementary interactions at energies much higher than man-made
accelerators can reach in foreseeable future. This opens opportunities to both measuring particle interactions (e.g.\ proton-nucleus, nucleus-nucleus, $\gamma$-nucleus, and $\nu$-nucleus interactions) at extreme energies as well as to probe fundamental physics, such as the smoothness of space-time or the validity of Lorentz invariance in yet unexplored domains \cite{Klinkhamer-07,Sigl-08}.

The energetic environments in which CRs are accelerated are likely to include matter or radiation fields with which the accelerated CR hadrons will interact, producing charged pions and kaons which then decay to neutrinos and photons. Neutrino emission would be an unambiguous signature for hadronic accelerators, since high energy photons can be produced also by inverse Compton scattering with energetic electrons. Moreover, neutrinos easily escape very dense local environments giving new insight into the most extreme cosmic objects ``hidden'' in any other observable. However, their detection at energies of several 100 TeV and above requires detection volumes of cubic kilometer scale. Besides serving in understanding the CR origin, neutrino telescopes can also be used to study neutrino properties and to search for new particles, such as magnetic monopoles, long living supersymmetric particles, or other exotica.

After decades of very slow progress because of lack of high statistics and high quality data, the chance to unravel the long mystery of CR origin has changed considerably. This is mostly due to the advent of high quality data from the Pierre Auger and the IceCube Observatories and of smaller scale experiments, such as the HiRes and Telescope Array Cosmic Ray and the ANTARES and Baikal-Lake Neutrino Observatories. This brief review summarizes recent results, some of which were presented at the last International Cosmic Ray Conference (ICRC) in Lodz, Poland.

\section{UHECRs: At the Doorway to Astronomy}

The 3000\, km$^2$ large Pierre Auger Hybrid Observatory \cite{Auger-04} is in full operation since summer 2008 and has already collected more than 17\,000~km$^2$sr\,yr of data at $E>10^{19}$~eV. This is about a factor of 2 more than HiRes \cite{Sokolsky-07} could collect in 10 years, and is a factor of 10 more than AGASA \cite{Shinozaki-06} could collect in about 20 years of operation. Already in 2007 the Pierre Auger Collaboration reported the observation of directional correlations of the most energetic CRs with the position of nearby AGN. Out of 27 events with energies above 57 EeV ($5.7\cdot10^{19}$\,eV) 20 were found to correlate within $3.2^\circ$ with AGN at redshift $z<0.018$ or $d<71$\,Mpc \cite{Abraham-07,AGN-long-08}. In case of an isotropic distribution, only 5.6 events were expected to correlate. However, this event sample included those data which were used to optimize the correlation parameters. Taking only independent data after the prescription, the corresponding numbers were 8 correlated events out of 12 observed ones. This was the first signature of anisotropy at the highest energies and by its directional structure it is also the first signature of extragalactic origin of the highest energy CRs. In the meantime, the data sample has doubled but the significance of the correlation signal has not become any stronger. It is still just above the 99\,\% C.L.\ because the correlation strength (fraction of correlated events) has decreased from about 70\,\% to 40\,\% \cite{AGN-ICRC09}. Statistical tests show, that this apparent change of the correlation signal is still in agreement with fluctuations. More sophisticated Likelihood tests in which the directional pattern of the 58 highest energy events is compared e.g.\ with a density map based on the recent Swift-BAT AGN catalogue shows that the data agree very well to the AGN map and that the chance probability to observe this in case of an isotropic event distribution is less than $10^{-5}$ \cite{AGN-ICRC09}. This underpins the anisotropy of the arrival directions and the overall correlation with the nearby cosmic matter distribution. On the other hand, HiRes - located in the Northern Hemisphere - analyzed their 13 highest energy stereo events with $E>56$~EeV and could not confirm the correlation signal of Auger \cite{Abbasi-AGN-08}. A comparison of the CR flux spectrum, however, shows that the Auger and HiRes energy scales differ by approx.\ 25\,\% \cite{Kampert-TAUP} so that the energy threshold applied by the HiRes Collaboration compares to about 44 EeV at the Auger energy scale. At this low threshold the Auger collaboration does not see any significant correlation, either. Also, the statistics of the Telescope Array (TA), which covers an area of 700 km$^2$ and became fully operational last year, is (and probably will remain) too low to help solving this issue. At the ICRC, TA reported that 2 out of 3 events show a correlation to a nearby AGN when applying the Auger parameters \cite{Sagawa-09}. Hence, more statistics needs to be collected before being able to give a definite answer about the origin of the AGN correlation signal. Moreover, as pointed out by Takami \cite{Takami-09}, directional correlations at the 3-5$^\circ$ scale are well possible in the south but could easily be destroyed in the Northern Hemisphere due to the strength and structure of the galactic magnetic field.

\begin{figure}[t]
\centerline{\includegraphics[width=0.8\textwidth]{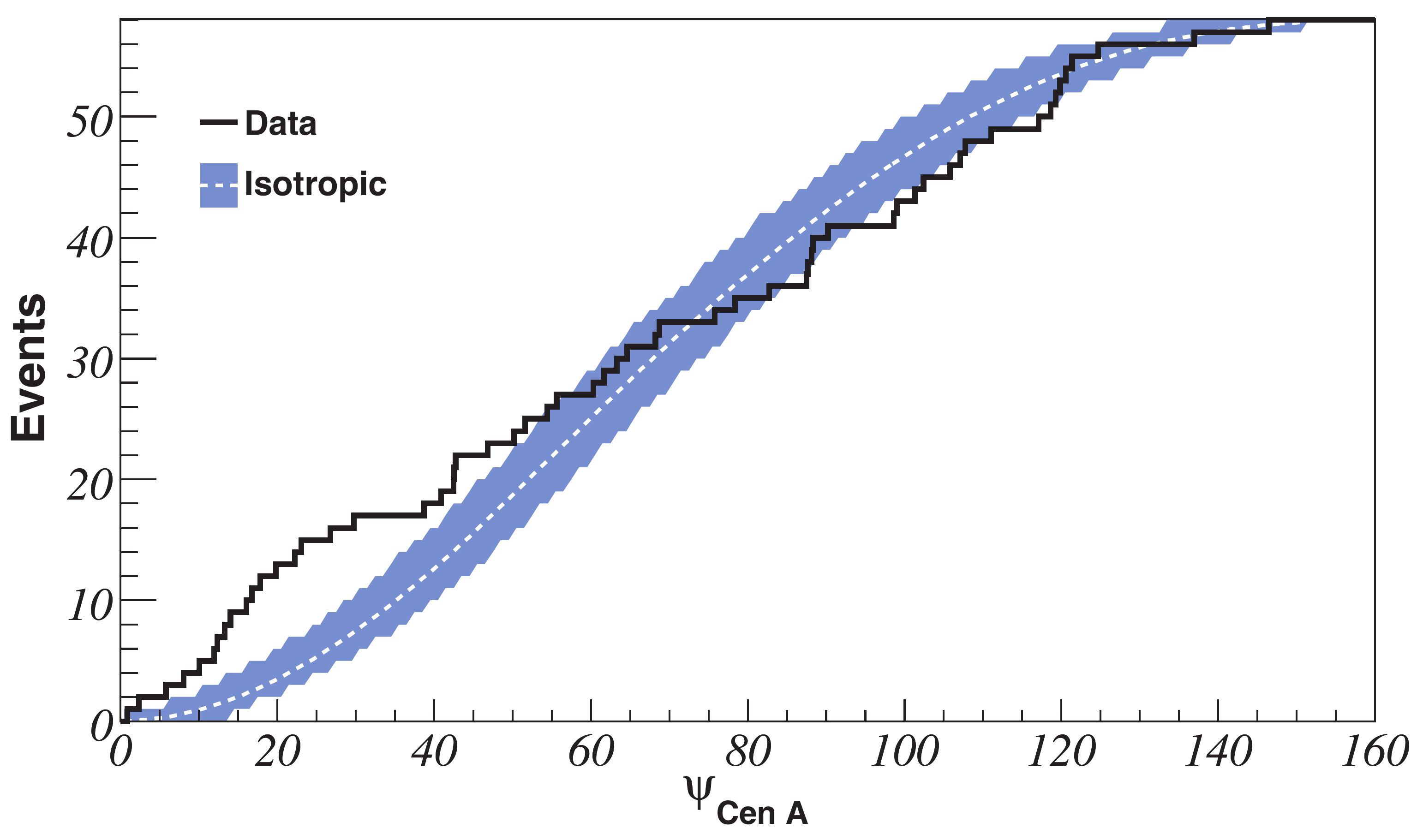}}
\caption[xx]{The cumulative number of events seen by the Pierre Auger Observatory with $E \ge 55$ EeV as a function of angular distance from Cen-A. The average isotropic expectation with approximate 68\,\% confidence intervals is shaded blue \cite{AGN-ICRC09}.}
\label{fig:cena-E}
\end{figure}

\begin{figure}[t]
\centerline{\includegraphics[width=0.8\textwidth]{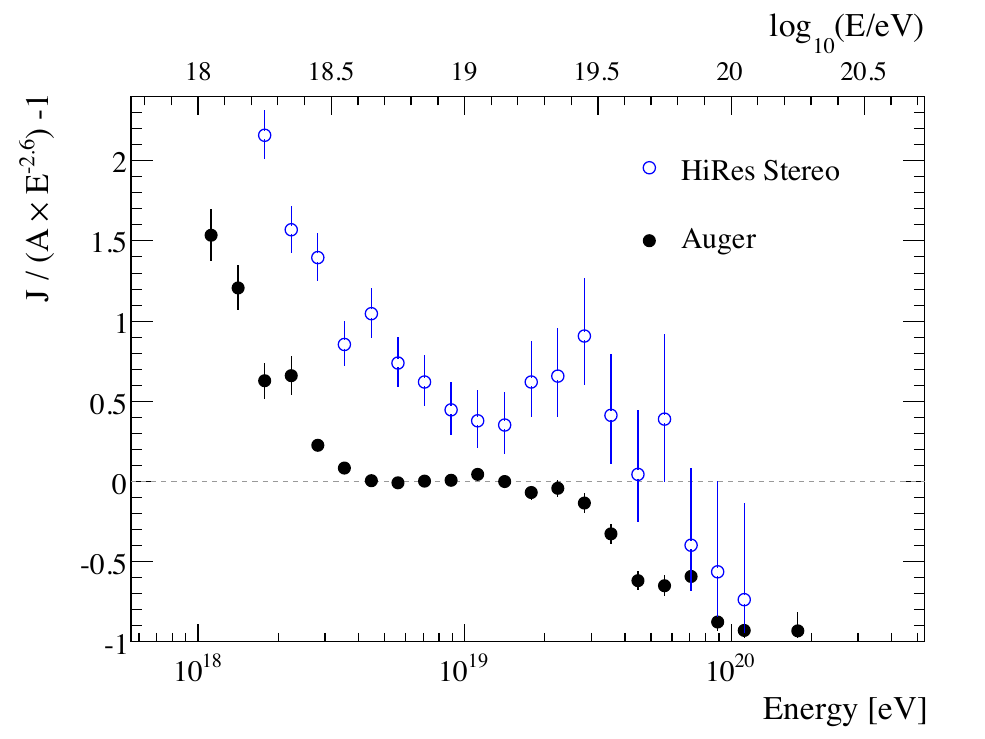}}
\caption[xx]{Fractional difference between energy spectrum of the Pierre Auger Observatory \cite{E-ICRC09} and a spectrum with an index of 2.6 compared to data from the HiRes \cite{Abbasi-stereo-09} instrument.}
\label{fig:Espec}
\end{figure}

Amongst all AGN, Centaurus A appears to be a highly interesting candidate. A large fraction of the high energy Auger events points towards the direction of this very extended source, as can be seen in Fig.\ \ref{fig:cena-E} \cite{AGN-ICRC09}. In a Kolmogorov-Smirnov test 2\,\% of isotropic realizations have a maximum departure from the isotropic expectation greater than or equal to the maximum departure for the observed events. The excess of events in circular windows around Cen-A with the smallest isotropic chance probability corresponds to a radius of $18^\circ$, which contains 12 events where 2.7 are expected on average if the flux were isotropic. It is worthwhile to mention that both HESS and Fermi-LAT confirmed Cen-A as an interesting source as they both reported the observation of high energy gamma rays from Cen-A. By contrast, the region around the Virgo cluster is densely populated with galaxies but does not have an excess of CR events above isotropic expectations.

Another very important step towards unveiling the origin of the sources of UHECR is provided by measurements of the CR energy spectrum (see Fig.\,\ref{fig:Espec}). Both Auger \cite{E-ICRC09,Abraham-08} and HiRes \cite{Abbasi-stereo-09,Abbasi-08} observe a distinct break in the energy spectrum at $E\simeq 40$~EeV. To enhance the visibility of the spectral shape, the fractional difference of the measured flux with respect to a reference flux $\propto E^{-2.6}$ is shown. The suppression of the flux at the highest energies and the ankle at $E\simeq 4\cdot 10^{18}$\,eV in the Auger data are evident. The spectral feature at the highest energies is in perfect agreement with expectations from the GZK-effect. Of course, sources running out of power could exhibit a similar feature. However, this would be a strange coincidence and in fact the onset of directional correlations observed by the Pierre Auger Observatory just above break energy supports the picture of the GZK-effect acting as a filter to nearby sources.

\section{High Energy Neutrinos: Strong Bounds on Astrophysical Models}

\begin{figure}[t]
\centerline{\includegraphics[width=0.8\textwidth]{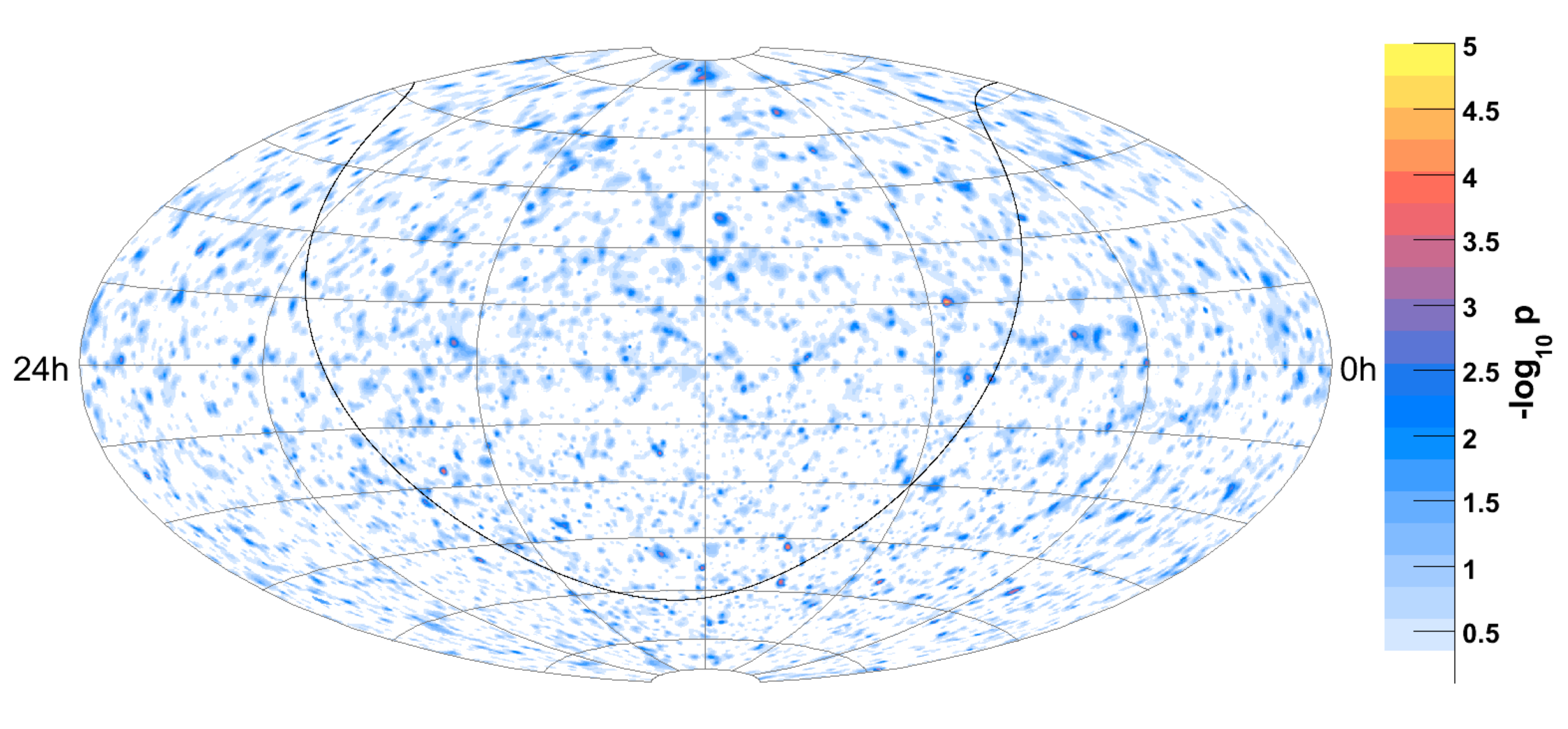}}
\caption[xx]{The map in equatorial coordinates shows the probability for a point source of high-energy neutrinos on the atmospheric neutrino background. It was obtained by operating IceCube with 40 strings for half a year \cite{Karle-ICRC, Spiering-EPS}. The ``hottest spot'' in the map represents an excess of 7 events which is not significant when taking into account the trial factors. The galactic plane is shown by the black line.}
\label{fig:neutrino-sky}
\end{figure}

Large volume detectors for astrophysical neutrinos were pioneered by the Baikal and AMAN\-DA telescopes in the Northern and Southern Hemisphere, respectively. Both have been successfully operated for about 10 years. In the Northern Hemisphere, European activities have now been concentrated in the Mediterranean sea with ANTARES near Toulon, NESTOR near Pylos in Greece, and NEMO off the eastern coast of Sicily.  The most advanced of these three projects is ANTARES comprising 12 operating strings with a total of about 900 photomultipliers. The three collaborations have now formed the {\sc KM3NeT} consortium with the goal to design and locate the future Mediterranean km-scale project. At the Geographic South Pole, AMANDA is now succeeded by the IceCube telescope. Both projects have a strong European involvement, too. IceCube has already deployed 59 of a total of 86 strings and is expected to be fully operational in February 2011. Operating a neutrino telescope in deep water and ice can be considered complementary as they face different technical challenges and need to account for the different properties of the optical media. Owing the largely different volumes and operation periods of the different neutrino telescopes, IceCube and AMANDA have provided the best presently existing bounds on astrophysical neutrinos and on searches for dark matter and other exotica.

As an example, we show in Fig.\,\ref{fig:neutrino-sky} the first full sky map of 6 months of IceCube 2008 data, based on 40 strings \cite{Karle-ICRC,Spiering-EPS}. This is the first result obtained with half of IceCube instrumented. To explore the full sky with one instrument only, the analysis made use of suppressing atmospheric downgoing muons through energy-sensitive cuts. The ``hottest spot'' in the map represents an excess of 7 events, an excursion from the atmospheric background with a probability of $10^{-4.4}$. After taking into account trial factors, the probability for this event to happen anywhere in the sky map is about 60\,\%. The background consists of 6796 neutrinos in the Northern Hemisphere and 10981 downgoing muons in the Southern Hemisphere. The energy threshold for the Southern Hemisphere increases with increasing elevation to reject the muon background by up to a factor of $\sim 10^{-5}$. The energy of accepted downgoing muons is typically above 100 TeV. The corresponding sensitivity to point sources will reach a level of $E_\nu^2 \, dN/dE_\nu < 10^{-12}$\,TeV cm$^{-2}$ s$^{-1}$ for 1 year of full IceCube and $5\cdot10^{-11}$\,TeV cm$^{-2}$ s$^{-1}$ for one year of Antares. 

\begin{figure}[t]
\centerline{\includegraphics[width=0.8\textwidth]{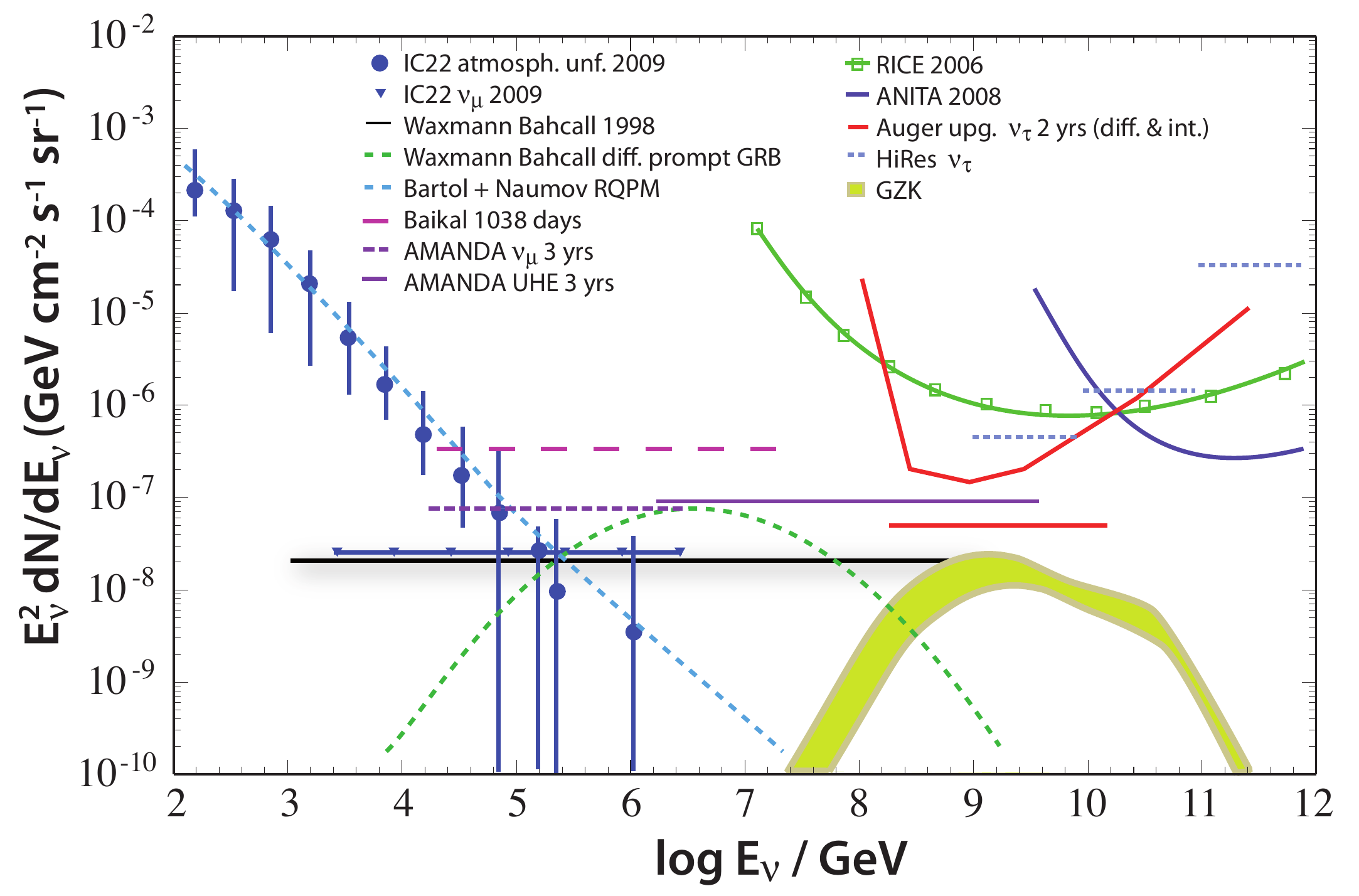}}
\caption[xx]{Measured atmospheric neutrino fluxes and compilation of latest limits on diffuse neutrino fluxes compared to predicted fluxes.}
\label{fig:nu-limits}
\end{figure}

Diffuse fluxes of neutrinos could arise for example from a class of sources too dim to be resolved individually. Assuming a non-thermal component, their flux is expected to have a harder energy spectrum than the atmospheric muon and neutrino backgrounds. Fig.\,\ref{fig:nu-limits} shows a preliminary measurement of the atmospheric neutrino flux obtained by IceCube instrumented with 22 strings \cite{Chirkin-09}. Thus, applying sufficiently high energy thresholds, extraterrestrial fluxes could either be detected or upper limits be placed. Moreover, at the highest energies $E_\nu \ga 10^{18}$\,eV, `cosmogenic neutrinos' from photo-pion production would be a guaranteed source if the GZK-effect exists. Searches for diffuse neutrino fluxes were performed by a large number of experiments operating at different energy regions and a compilation of recent data (partly presented or updated at ICRC09 \cite{AGN-ICRC09,Abraham-neutrino-09,Mase-09}) is presented in Fig.\,\ref{fig:nu-limits}. As can be seen, the current limits are approaching both the Waxmann-Bahcall and the cosmogenic flux (labelled GZK in Fig.\, \ref{fig:nu-limits}) predictions. Assuming that AGN or GRBs radiate similar energies in photons and CRs, fluxes of high energy neutrinos from GRBs and AGN scenarios can be estimated \cite{Halzen-09} to be at a level $E_\nu^2 \, dN/dE_\nu \approx 5 \cdot 10^{-8}$\,TeV cm$^{-2}$ s$^{-1}$ sr$^{-1} \times x_\nu$, with $x_\nu \sim 0.05$. Hence, it needs at least a cubic kilometer scale telescope to enter the discovery domain. Similarly, it will take several years of Auger-scale experiments to detect the cosmogenic neutrino fluxes at about $10^9$~GeV, depending on the spatial source distribution and composition of CRs.

\section{Discussion and Outlook}

Remarkable progress has been made in high energy cosmic ray and neutrino physics over the last two years. In case of UHECRs, the GZK-structure in the energy spectrum is observed with high statistical accuracy and anisotropies at the highest energies have emerged from the data. The coincidence of seeing anisotropies just above the GZK-threshold confirms the picture of the GZK-effect acting as a filter to nearby sources. Even though the directional correlation to nearby AGN is still under debate, the small scale of the angular correlations suggests light particles as primary particles. However, observations of the position of the shower maximum, $X_{\rm max}$, in the atmosphere and the level of fluctuations of $X_{\rm max}$ suggest mixed heavy component \cite{M-ICRC09}. This puzzle is to be solved. It could be due to the poorly known hadronic interactions at cms energies 2-3 orders of magnitude higher than at the future LHC collider and/or due to a strongly increasing cross section at the highest energies, or due to weaker magnetic fields than generally assumed. 

The sensitivity of the neutrino telescopes to point sources and diffuse fluxes has been improved over several orders of magnitude and over a wide range of energies during the last years. Still, no point sources or diffuse fluxes are observed. About 10 months of full IceCube equivalent data are at hand now and the chances are high to discover astrophysical neutrino sources within the next few years. However, based on the information from TeV gamma and UHECR observations, the neutrino fluxes from galactic and extragalactic sources seem to be lower than thought a few years ago. Hence, analyzing individual neutrino sources in some detail at TeV energies will require instruments much larger than a cubic kilometer. This should be considered when designing the future {\sc KM3NeT} telescope. Similarly, due to the GZK break in the UHECR energy spectrum, much larger observatories than available at present are required to collect sufficient statistics for CR-astronomy of individual sources. Related R\&D efforts, such as radio and acoustic particle detection methods, are ongoing to face these challenges.

\subsection*{Acknowledgement}
I would like to thank the organizers of the EPS high energy physics conference for giving me the opportunity to present these results at such an important meeting. Also, it is a pleasure to thank my colleagues from Auger and IceCube for stimulating discussions. The German Ministry for Research and Education (BMBF) is gratefully acknowledged for financial support.


\end{document}